\documentclass{mem}
\usepackage{natbib}\usepackage{txfonts}\usepackage{balance}
\usepackage{graphicx}
\usepackage{txfonts}
\idline{75}{282}
\begin{document}

\def\lsun{L$_\odot$}
\def\msun{M$_\odot$}
\def\mstar{$M_\star$}
\def\mgas{$M_{gas}$}
\def\fgas{$f_{gas}$}
\def\zsun{Z$_\odot$}
\def\mc{\multicolumn}
\def\cgs{erg~cm$^{-2}$sec$^{-1}$}
\def\mic{$\mu$m}
\def\arcsec{$^{\prime\prime}$}
\def\yeff{$y_{eff}$}
\def\ysun{$y_\odot$}

\def\ha{H$\alpha$}
\def\hb{H$\beta$}
\def\oii{[OII]$\lambda$3727}
\def\nii{[NII]$\lambda$6584}
\def\neiii{[NeIII]$\lambda$3869}
\def\oiiia{[OIII]$\lambda$4958}
\def\oiiib{[OIII]$\lambda$5007}
\def\oiii{[OIII]$\lambda$4958,5007}
\def\oiit{[OII]\small{3727}}
\def\neiiit{[NeIII]\small{3869}}
\def\oiiiat{[OIII]\small{4958}}
\def\oiiibt{[OIII]\small{5007}}
\def\hei{HeI\small{3889}}

\def\aj{AJ}
\def\apj{ApJ}
\def\apjs{ApJS}
\def\apjl{ApJL}
\def\aap{A\&A}	
\def\mnras{MNRAS}
\def\pasp{PASP}
\def\araa{ARAA}
\def\nat{Nature}

\title{
The metallicity properties of long-GRB hosts
}

   \subtitle{}

\author{
F. \,Mannucci\inst{1,2} 
          }

  \offprints{F. Mannucci}

\institute{
Istituto Nazionale di Astrofisica --
Osservatorio Astrofisico di Arcetri, 
Largo E. Fermi 5 I-50125 Florence, Italy
\email{filippo@arcetri.astro.it}
\and
Harvard-Smithsonian Center for Astrophysics,
60 Garden Street, 
Cambridge, MA 02138
USA 
}

\authorrunning{Mannucci}

\titlerunning{Metallicity of LGRB hosts}

\abstract{
The recently-discovered Fundamental Metallicity Relation (FMR), which is
the tight dependence of metallicity on both mass and SFR, proves to be a very useful 
tool to study the metallicity properties of various classes of galaxies. 
We have used the FMR to study the galaxies hosting long-GRBs.
While the GRB hosts have lower metallicities than typical galaxies 
of the same mass, i.e., they are below the mass-metallicity relation, 
they are fully consistent with the FMR. 
This shows that the difference with the mass-metallicity 
relation is due to higher than average SFRs, and that 
GRBs  with optical afterglows do not preferentially select 
low-metallicity hosts among the star-forming galaxies. 
}
\maketitle{}

\begin{figure}
\resizebox{\hsize}{!}{\includegraphics[clip=true]{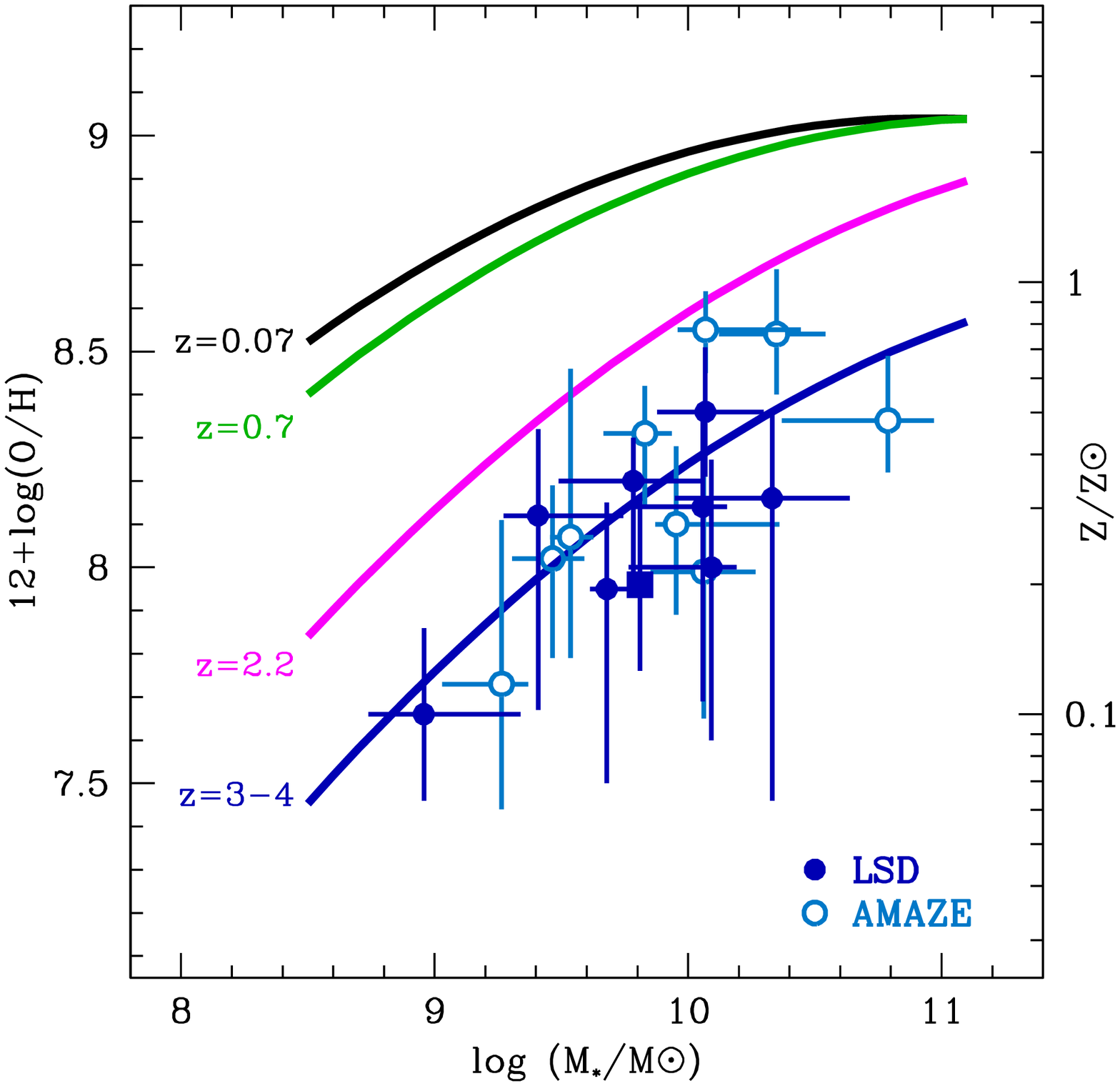}}
\caption{\footnotesize
Evolution of the mass-metallicity relation from local to high redshift galaxies
from \cite{Mannucci09b}.
Data are from \cite{Kewley08} (z=0.07), \cite{Savaglio05} (z=0.7), 
\cite{Erb06a} (z=2.2) and \cite{Mannucci09b} (z=3--4).
}
\label{fig:massmetevol}
\end{figure}

\section{Introduction}

The gas-phase chemical abundance in galaxies is influenced by several 
effects: star formation and evolution, which reduce the amount of gas 
and increase the amount of metals, 
infall of metal-poor gas from the outer part of the galaxy and from the intergalactic medium, 
and outflow of enriched material due to feedback from SNe and AGNs. 
As a consequence, gas-phase metallicity is a fundamental test for all models of 
galaxy formation.

A fundamental discovery has been the relation between stellar mass \mstar\ 
(or luminosity) and metallicity 
\citep{McClure68,Lequeux79,Garnett02,Lamareille04,Pilyugin04,Tremonti04,Lee06,
Liang06,Pilyugin07},
with more massive galaxies showing higher metallicities.
The origin of this relation is debated, and many different explanations
have been proposed, including ejection of metal-enriched gas
(e.g., \citealt{Edmunds90,Tremonti04}),
``downsizing'', which is a systematic dependence of the efficiency of star 
formation on galaxy mass  
(e.g., \citealt{Brooks07,Mouhcine08,Calura09a}), variation of the IMF
with galaxy mass \citep{Koppen07}, and infall of metal-poor gas
\citep{Finlator08,Dave10}.

The evolution of the luminosity-metallicity and mass-metallicity relations 
has been studied by many authors at z$<$1.5 
\citep{Contini02,Kobulnicky03,Maier04,Liang04,Kobulnicky04,Maier05,Savaglio05,
Shapley05a,Lee06,Lamareille06,Maier06,Liu08,Cowie08,Rodrigues08,Hayashi09a,
Lara-Lopez09a,Lamareille09a,Perez-Montero09,Queyrel09,Vale-Asari09,Thuan10,Zahid11}
at z$\sim$2--3 \citep{Erb06a,Hayashi09a,Yoshikawa10,Richard11}, 
and at z$>$3
\citep{Pettini01,Maiolino08,Mannucci09b,Lemoine-Busserolle10}, finding 
a strong and monotonic evolution, with metallicity decreasing with redshift
at a given mass (see Fig.\ref{fig:massmetevol}).
Both the shape and the normalization of the mass-metallicity relation are sensitive
to the metallicity calibration used \citep{Kewley08,Peeples11}, which can differ 
significantly.
Part of the differences is due to the secondary nature of nitrogen, whose abundance ratio 
with oxygen is expected and observed to vary during the galaxy lifetime. As several metallicity 
calibrations are based on the flux ratio of oxygen emission lines to the [NII]$\lambda$6584 line, 
this uncertainty also affects the oxygen abundance
(e.g., \citealt{Pilyugin04,van-Zee06,Liang06,Perez-Montero09b,Queyrel09,Lopez-Sanchez10a,Pilyugin11b,Thuan10}). 
This point will be addressed in a forthcoming paper (Maiolino et al., in preparation).
Despite these problems, the evidence of evolution of the mass-metallicity relation 
is not affected by calibration uncertainties and is a very solid result.

Some authors \citep{Erb06a,Erb08,Mannucci09b} 
have studied the relation between metallicity and gas fraction, i.e., 
the effective yields. These results can be explained as a consequences of 
infall in high redshift galaxies. 
Also, \cite{Cresci10} studied the metallicity maps of three star-forming 
galaxies at z$>$3, founding regions of low metallicity associated with the peak
of star-formation. This evidence can be explained assuming that some metal-poor 
infalling gas both fuels star formation and dilutes metallicity.
If infall is at the origin of the star formation activity,
and outflows are produced by exploding supernovae (SNe),
a relation between metallicity and SFR
is likely to exist. In other words, SFR is a parameter
that should be considered in the scaling relations that include metallicity.

\begin{figure*}
\resizebox{\hsize}{!}{
   \includegraphics[width=0.48\textwidth]{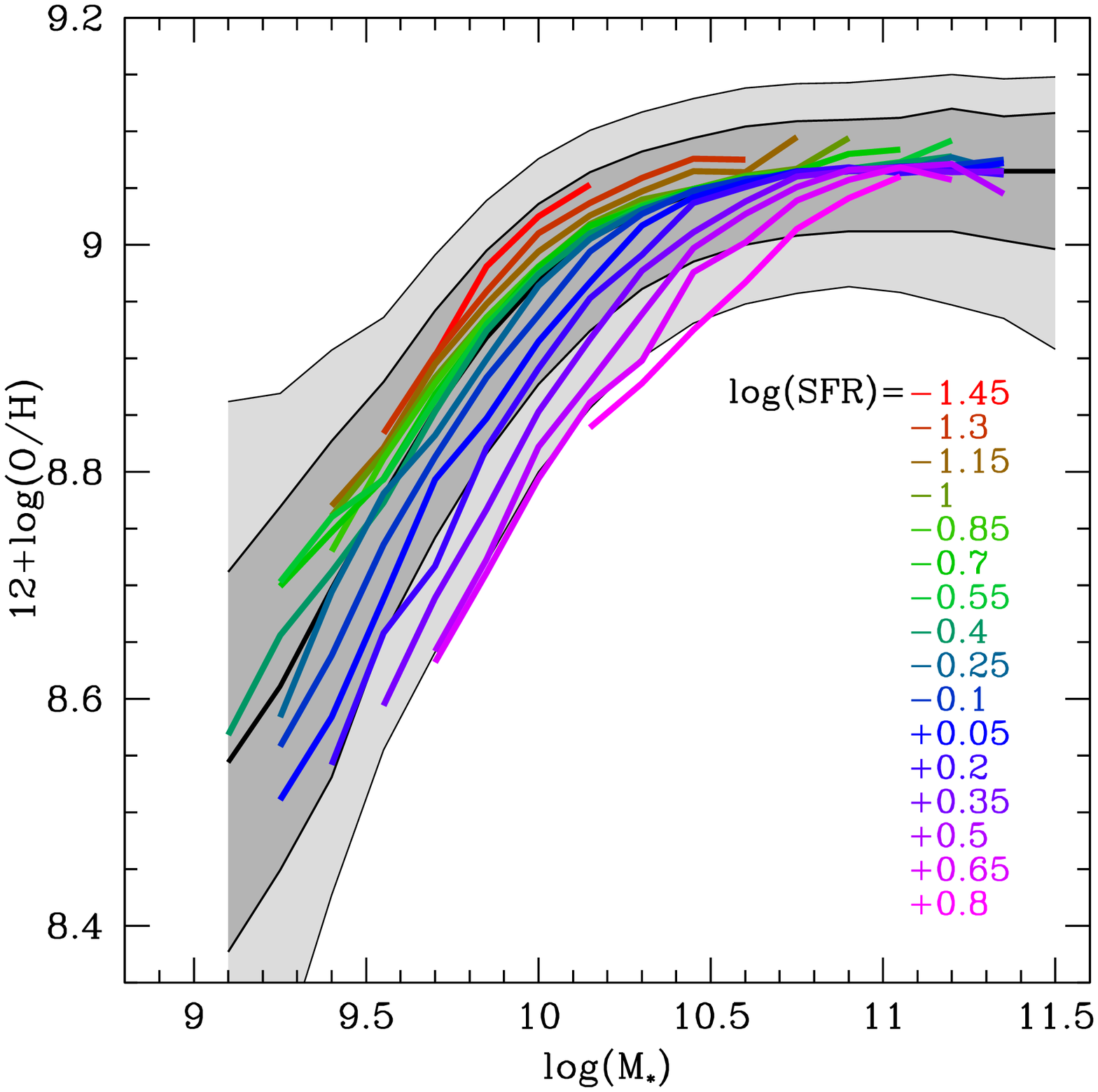} 
   \includegraphics[width=0.48\textwidth]{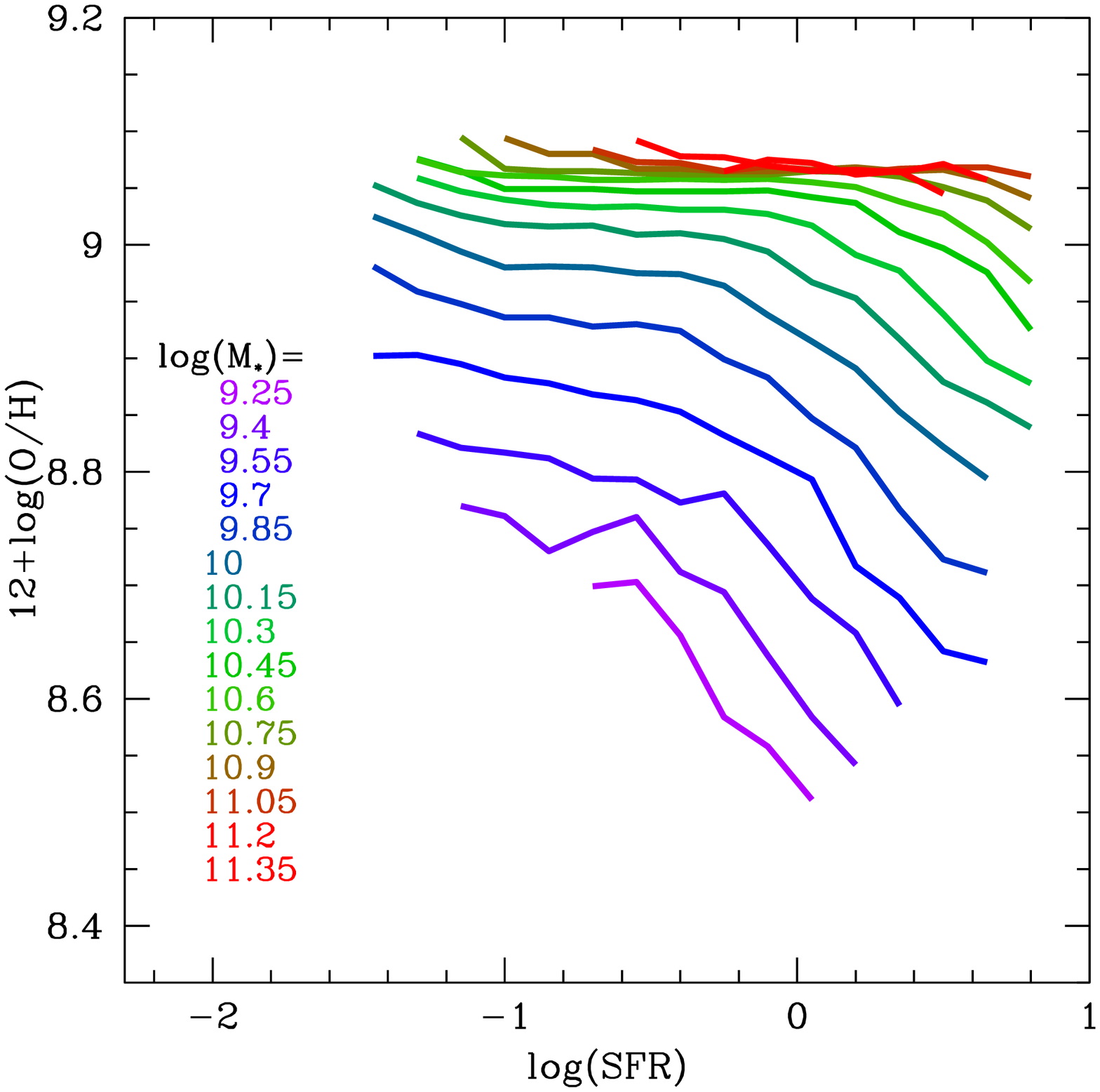} 
}
\caption{\footnotesize
{\em Left panel:} The mass-metallicity relation of local SDSS galaxies. 
The grey-shaded areas contain 64\% and 90\% of all SDSS galaxies, with the 
thick central line showing the median relation. 
This is very similar to what has been found by \cite{Tremonti04}, but 
the dispersion of our sample is somewhat smaller, 0.08dex instead of 0.1dex.
The colored lines show the 
median metallicities, as a function of \mstar, 
of SDSS galaxies with different values of SFR. Metallicity shows a systematic decreas
with increasing SFR.
{\em Right panel:} median metallicity as a function of SFR for galaxies of 
different \mstar.  At all \mstar\ with log(\mstar)$<$10.7,
metallicity decreases with increasing SFR 
at constant mass .
}
\label{fig:massmet}
\end{figure*}

\begin{figure}[t]
\resizebox{\hsize}{!}{
	\includegraphics[width=\textwidth]{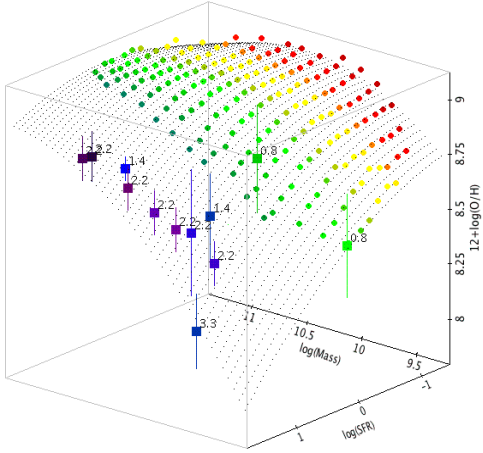} 
}
\caption{\footnotesize
The Fundamental Metallicity Relation  plotted in the 3D space defined by
\mstar, SFR and gas-phase metallicity. 
Circles without error bars are the median values of metallicity of local SDSS galaxies
in bin of \mstar\ and SFR, color-coded with SFR.
The black dots show a second-order fit to these SDSS data, 
extrapolated toward higher SFR.
Square dots with error bars are the median values of high redshift galaxies, 
as explained in the text.
Labels show the corresponding redshifts.
All the the high-redshift data,
except the point at z=3.3, are found on the same surface defined by low-redshift data.
}
\label{fig:3D}
\end{figure}
  
\begin{figure}
\resizebox{\hsize}{!}{\includegraphics[]{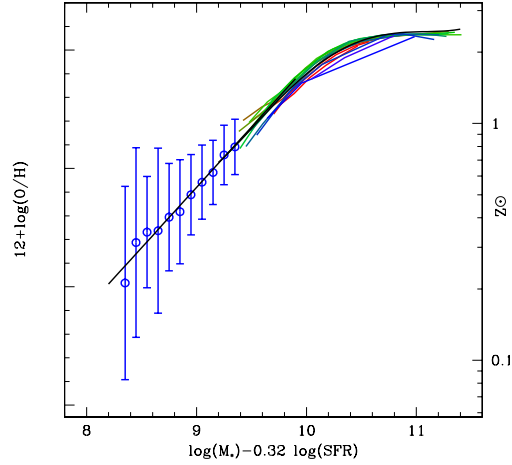}}
\caption{\footnotesize
Extension of the FMR towards low-mass galaxies, adapted from Mannucci et al. (2011). 
The blue dots show the median 
metallicity of low mass SDSS galaxies in bins of
$\mu_{0.32}$
in solar units, as defined in Mannucci et al. (2010). 
The coloured lines are local SDSS galaxies from Mannucci et al. (2010), color-coded
from red to blue according to increasing SFR. 
The black thick line shows the polynomial fit.
For comparison, the black dotted line is the extrapolation of the 2nd degree fit to 
the FMR of the SDSS galaxies as defined in Mannucci et al. (2010) and plotted for SFR=0. 
\label{fig:lowmass}
}
\end{figure}

\begin{figure}
\resizebox{\hsize}{!}{
	\includegraphics[width=0.48\textwidth]{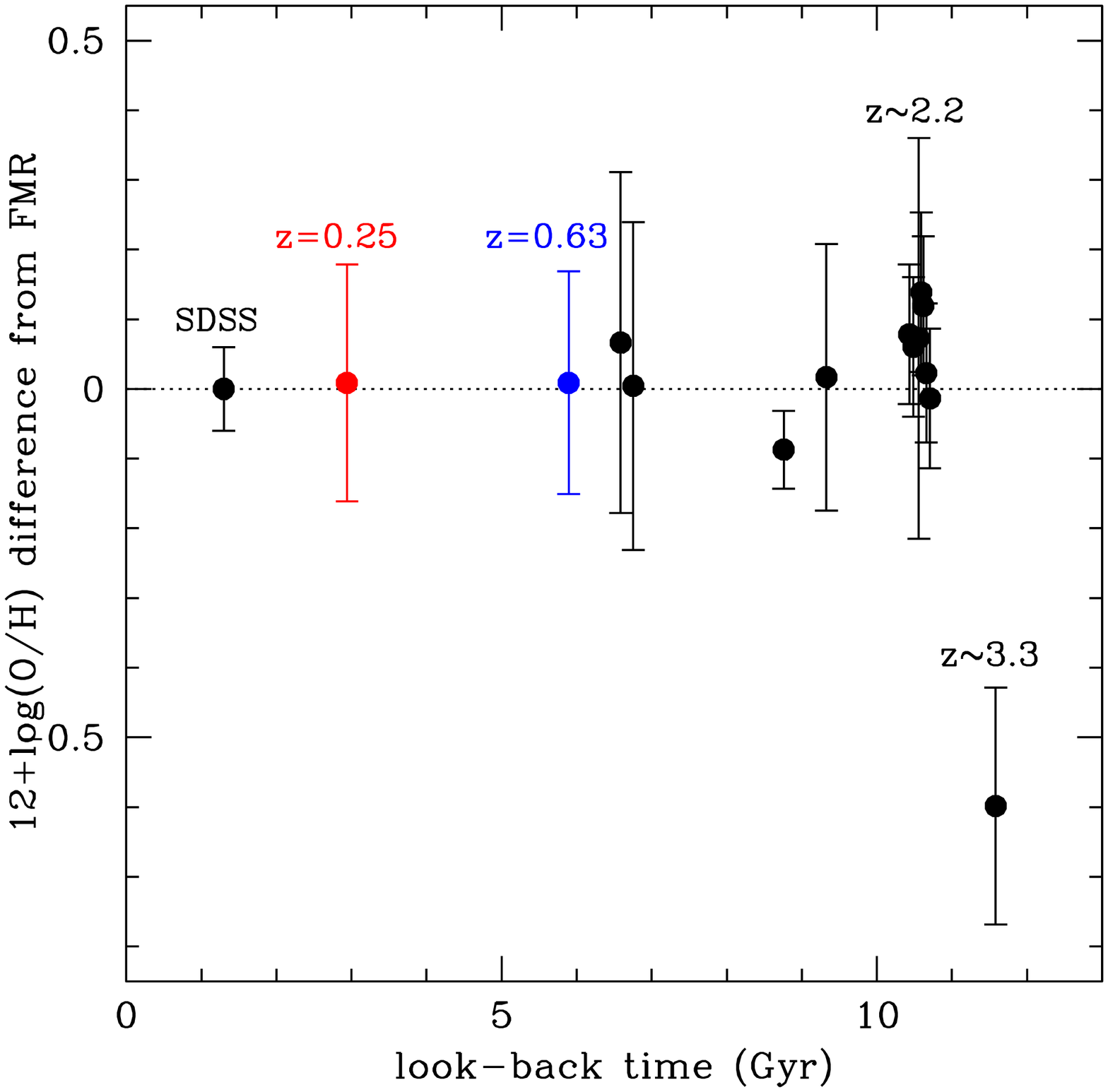} 
}
\caption{\footnotesize
Metallicity difference from the FMR for samples of galaxies at different redshifts,
from Cresci et al. (in preparation). The black dots are the original data from \cite{Mannucci10}, 
the colored dots are zCosmos galaxies.
The SDSS galaxies defining the relation are showing at z$\sim$0.1 with their
dispersion around the FMR. All the galaxy samples up to z=2.5 are consistent 
with no evolution of the FMR defined locally. 
Metallicities lower by $\sim$0.6~dex are observed at z$\sim$3.3.
}
\label{fig:plotevol}
\end{figure}

\section{A Fundamental Metallicity Relation in the local universe}

In \cite{Mannucci10} we studied the dependence of metallicity on both 
mass and SFR in SDSS galaxies. As shown in Fig.\ref{fig:massmet},
at constant stellar mass, metallicity anti-correlates with SFR, i.e.,
galaxies with higher SFRs also show lower metallicities.
The dependence of metallicity on \mstar\ and SFR can be better visualized 
in a 3D space with these three coordinates, as shown in Fig.~\ref{fig:3D}.
In this space, SDSS galaxies appear to 
define a tight surface, named the Fundamental Metallicity Relation (FMR).
The introduction of the FMR results in a significant reduction of residual 
metallicity scatter with respect to the simple mass-metallicity relation. 
The dispersion of individual SDSS galaxies around the FMR is about 
$\sim$0.05~dex i.e, about 12\%, and this 
scatter is consistent
with the intrinsic uncertainties in the measures of metallicity , mass, and SFR.

\cite{Mannucci11a} have extended this relation toward lower stellar masses, down to 
about $10^8$\msun (see Fig.\ref{fig:lowmass}).

The resulting FMR can be expressed by:
\begin{equation}
\begin{array}{rl}
12&+log(O/H)=                    \\
   &=8.90+0.37m-0.14s-0.19m^2                    \\
   &~~~+0.12ms-0.054s^2~~~~~~~~\rm{for}~\mu_{0.32}\ge9.5 \\
   &=8.93+0.51(\mu_{0.32}-10)~~~\rm{for}~\mu_{0.32}<9.5 \\
\end{array}
\end{equation}
where $m$=log(\mstar)-10 and $s$=log(SFR) are in solar units,
$\mu_{0.32}=log(M_\star)-0.32*log(SFR)$ , and
$8<log(M_\star)<11.5$ and $-1.5<log(SFR)<1.0$.

\smallskip
It is interesting to compare the metallicity properties of high redshift galaxies 
to the local FMR. In \cite{Mannucci10} we did this computation using data from the literature 
and our sample of galaxies z$>$3. Astonishingly, we found no evolution
up to z=2.5, i.e., high redshift, star-forming galaxies follow the same FMR defined 
by local SDSS galaxies, with no sign of evolution.  
This has been confirmed by Cresci et al (in preparation), who studied a large sample of galaxies at 
z$\sim$0.2 and z$\sim$0.6 from the zCosmos sample, obtaining no indication of evolution off the FMR.

This is an unexpected result, as simultaneously the
mass-metallicity relation is observed to evolve rapidly with redshift
(see Fig.\ref{fig:massmetevol}).
The solution of this apparent paradox is that distant galaxies have, on average,
larger SFRs, and, therefore, fall in different parts of the same FMR.
Several recent studies 
\citep{Richard11,Erb10,Nakajima11,Trump11,Contini11,Atek11}
have presented samples of high-redshift galaxies whose SFRs are significantly higher or 
lower than most of the previously-known galaxies.
In all these cases, a discrepancy with the mass-metallicity relation at
that redshift is observed, together with a good agreement with the FMR. 
In this respect, the FMR has a real predictive 
power, i.e., the metallicity of a star-forming galaxy can be predicted at the 0.1dex level 
from its mass and SFR.

This no evolution is observed up to z$\sim$2.5. 
Galaxies at z$\sim$3.3 show metallicities 
lower of about 0.6~dex with respect to
both the FMR defined by the SDSS sample and galaxies at 0.5$<$z$<$2.5. 
This is an indication that some evolution of the FMR 
appears at z$>$2.5, although its size 
can be affected by several potential biases (see \citealt{Mannucci10} for 
a full discussion). 
This result is confirmed by Sommariva et al. (in preparation) who studied
the relation between mass and {\it stellar} metallicity in star-forming galaxies at z$>$3. 
They found a good agreement between stellar and gas-phase metallicities, confirming the 
evolution of the FMR.

\smallskip
Several authors are now using the FMR as an additional constraint for their models of galaxy formations
and are proposing different ways to explain the existence of such a tight correlation
\citep{Mannucci10,Campisi11,Dib11,Peeples11,Dave11c,Dave11d}.
The actual physical significance of the FMR is not yet completely clear, but its explanation will probably 
involve some complex interaction between infalling metal-poor gas, outflowing metal-enriched gas,
mass- and metallicity-dependent efficiency of star formation, and possibly systematic differences in the
IMF.
\section{The metallicity of GRB hosts}

Recent studies on the final evolutionary stages of
massive stars \citep{Woosley06,Fryer99} have suggested that a Wolf-Rayet star can
produce a long GRB if its mass loss rate is low, which is
possible only if the metallicity of the star is lower than $\sim
0.1-0.3\,Z_{\odot}$. 
In this view, GRBs may occur preferentially
in galaxies with low-metallicity \citep{Fynbo03,Prochaska04,Fruchter06},
even if low-metallicity progenitors can also be present in hosts with relatively high 
metallicities \citep{Campisi09}.
 
Observationally, the role of metallicity in driving the GRB phenomena remains unclear and 
it is still debated \citep{Fynbo03,Prochaska04,Fynbo06a,
Wolf07,Price07,Modjaz08,
Kocevski09,Savaglio09,Graham09a,Graham09c,Levesque10b,Levesque10c,Levesque10e,
Svensson10,Fan10b}.
Many recent studies have compared GRB hosts to the rest of the galaxies
(see, for example, \citealt{Fynbo08}). In particular, some of these studies
have compared the observed mass-metallicity relation (or luminosity-metallicity
relation) of the two populations, finding that 
long GRB host galaxies fall below the relation for
the normal galaxy population, i.e., GRB hosts are less enriched 
in metals than the typical 
galaxies of the same stellar mass.

\begin{figure*}
\resizebox{0.95\hsize}{!}{
	\includegraphics[angle=-90]{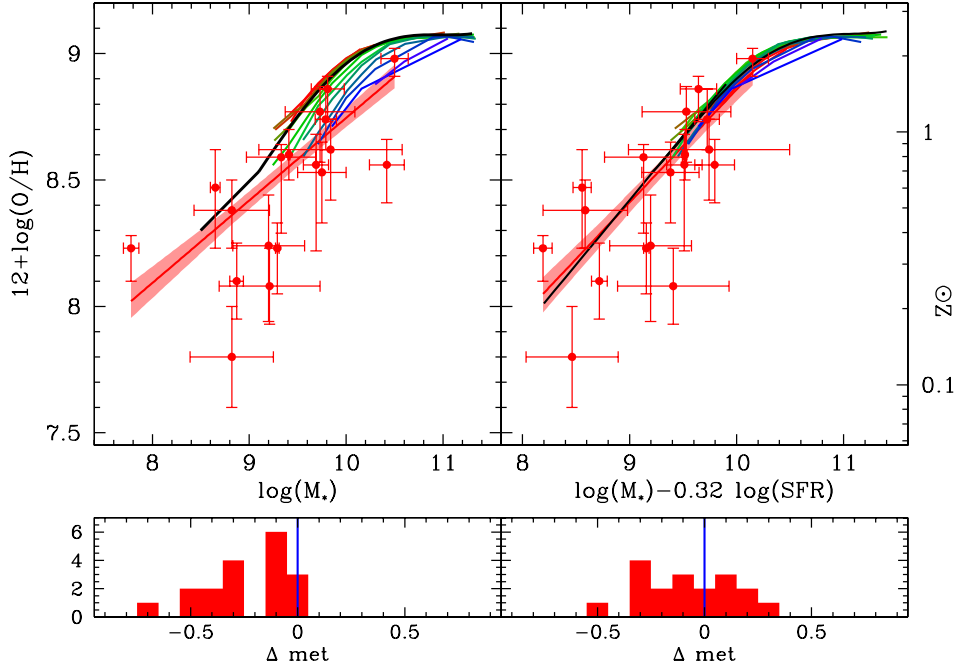} 
}
\caption{\footnotesize
Long-GRB hosts (red dots) are compared with the Mass-Metallicity relation ({\em left}) and the FMR ({\em right})
of local galaxies. Symbols are as in Fig.\ref{fig:lowmass}. 
The red thick line is a linear fit to GRB host data,
with $\pm1\sigma$ bands shown in light red. 
The {\em lower panels} show the difference between 
the metallicity of the GRB hosts and mean relations of SDSS galaxies (black curves).
It is clear that GRB host follow a different mass-metallicity
relation and show systematically lower metallicities.  In contrast, they are in perfect agreement with the 
predictions of the FMR. In other words, they have the metallicity that is expected based of their mass and SFR.
\label{fig:grbmet}
}
\end{figure*}

In order to check whether this bias exists, we have considered the properties of the 
GRB hosts in the light of the observed FMR.
To this extent, we have collected all the GRB host 
galaxies at z$<$1 with available observations
to measure, at the same time, 
stellar mass, SFR, and gas phase metallicity.
Line fluxes of long GRB hosts have been published by several authors 
\citep{Savaglio09,Han10a,Levesque10d}. 
We have measured gas-phase metallicities by simultaneously considering all 
the flux ratios among the relevant emission lines, 
and using the calibration in \cite{Maiolino08}. 
Dust extinctions have been obtained using the Balmer decrement.
SFR have been estimated from \ha\ corrected for extinction, 
using the calibration in \cite{Kennicutt98}. 
Finally, stellar masses have been taken from  \cite{Savaglio09}.

These data are plotted in fig.~\ref{fig:grbmet} and compared with both
the mass-metallicity relation (left panel) and the FMR (right panel) of local SDSS galaxies.
The comparison with the mass-metallicity relation shows that,
as already obtained by \cite{Levesque10d} and \cite{Han10a}, GRB host galaxies have lower 
metallicity than galaxies of the same mass.
In contrast, we also find that GRB hosts
do follow the FMR and its extension towards low masses, without any significant discrepancy.
In other words, when the dependence on SFR is properly taken into account,
the metallicity properties of long GRB hosts do not differ substantially from
those of the typical field population. 

As a consequence, 
the typical low, sub-solar  metallicity found in many recent studies
does not mean that GRBs occur in special, low-metallicity galaxies, 
but rather it is a consequence of the well-known
link between the GRB event and the death of a massive stars, 
which produces a relation between long GRBs and star formation
\citep{Totani97,Porciani01}.
In the local universe, about 70\% of all star formation activity occurs in 
 galaxies with masses between $10^{9.5}$ and $10^{10.2}$\msun
\citep{Mobasher09}, where most of the GRB of our sample are also found. 
The low metallicities are a consequence of the low masses and of the high SFRs.
It should be noted that our sample consists mostly of long GRBs whose position has been provided 
by the detection of their optical afterglow. It is known that a population of 
GRBs with a bright X-ray afterglow and lacking of optical counterparts 
does exist, the so-called dark GRBs, and most of them
reside in dusty environments (e.g., \citealt{Perley09b,Kupcu-Yoldas10}).
It is possible that 
dark GRB hosts would populate the region of the FMR at high values of $\mu_\alpha$
and metallicity.

Our results have an important implication for future studies of the high redshift universe. 
Given that GRB hosts appear to be normal, actively star-forming galaxies, 
large samples of GRB hosts
can be used to study the FMR of normal starburst galaxies.
This is particularly exciting since GRBs may allow to
extend current studies of the FMR both toward low values of $\mu_\alpha$ and 
toward higher redshift, in principle up to 
extremely high redshifts, at least up to $z\sim 8$ as shown by GRB 090423
\citep{Salvaterra09}.

\bibliographystyle{/Users/filippo/arcetri/Papers/aa-package/bibtex/aa}
\bibliography{/Users/filippo/arcetri/bibdesk/Bibliography}
\end{document}